\documentclass[useAMS,usenatbib]{mn2e}
\usepackage{graphicx}

\def\la{\raise.5ex\hbox{$<$}\kern-.8em\lower 1mm\hbox{$\sim$}}
\def\ma{\raise.5ex\hbox{$>$}\kern-.8em\lower 1mm\hbox{$\sim$}}

\def\msol{M$_{\odot}$ }
\def\Lsol{L$_{\odot}$ }

\def\kms{$\rm km\, s^{-1}$}
\def\cm3{$\rm cm^{-3}$}
\def\Ts{$\rm T_{*}$~}
\def\Vs{$\rm V_{s}$~}
\def\n0{$\rm n_{0}$}
\def\B0{$\rm B_{0}$}

\def\erg{$\rm erg\, cm^{-2}\, s^{-1}$}
\def\mum{$\mu$m~}

\def\L12{L$_{12\mu m}$~}
\def\F12{F$_{12\mu m}$~}
\def\agr{a$_{gr}$}
\def\Hb{H${\beta}$}
\def\Ha{H${\alpha}$}

\def\Ly{Ly$\alpha$~}

\def\Tbb{T$_{bb}$}
\def\ff{{\it ff}}

\def\RO3{R$_{[OIII]}$}
\def\fs{f$_s$~}

\title[Modelling  line and continuum spectra of SGRB hosts]{Modelling line  ratios and   continuum SED of NGC4993 and 
 other short  GRB host galaxies 
}

\author[M. Contini]{M. Contini 
\\
School of Physics and Astronomy, Tel Aviv University, Tel Aviv
69978, Israel \\
}

\begin{document}


\pagerange{\pageref{firstpage}--\pageref{lastpage}} \pubyear{2009}

\maketitle

\label{firstpage}

\begin{abstract}

We present a detailed spectral modelling of NGC4993 -- the
host galaxy of GW170817 --  and other SGRB host galaxies. In order to
 determine their  physical conditions and the element abundances,
we have gathered spectroscopic and photometric data  from the literature.
The  observation data  are  sometimes   missing
preventing us from fully constraining the model.
However, for  most of  the SGRB hosts  the [OIII]5007/\Hb~ and [NII]6548/\Ha~ line ratios  are reported. 
 The analysis of NGC4993 by a composite model (photoionization+shock)
 confirms that an AGN, most probably a  LINER or a LLAGN, is the  gas photoionization source.
Shock velocities and preshock densities are similar to those found in the narrow line region of AGN. 
 O/H and N/H  have solar values. 
For the other SGRB of the sample we have found that O/H ratios are  nearly solar,
while N/H cover a much larger range  of values  at redshifts close to 0.4. 
In NGC4993, the relative contribution to the SED of an old stellar
population, characterized by a black body temperature of \Tbb=4000K, with
respect to Bremsstrahlung is higher by a factor $>$ 100 than in most of the
local AGN and SB galaxies. For the other SGRB composing the sample, \Tbb~
ranges between 2000K for SGRB100206A and 8000K for SGRB111117A.

\end{abstract}


\section{Introduction}

 Gamma-ray bursts (GRB) are short, intense and isolated flashes
(Berger 2013), peaking in the gamma-ray band and occurring at an average rate of one event per 
day over the whole sky (D'Avanzo 2015) 
with  different spectral and temporal properties. 
The current scenario distinguishes  GRBs in "short-hard and long-soft" (Berger 2013) bursts. 
Short duration GRB (SGRB) last less than 2s, while long GRB (LGRB) have longer duration
(Kouveliotou et al 1993).
Both the localization and detection of  afterglows and hosts and  the evidence of lower energy and density scales
suggest (Fox et al 2005, Barthelmy et al 2005, Berger et al 2005)
 that SGRB are cosmological in origin, with low energetic afterglow and
that their  progenitors are not massive stars (Berger 2013). 
Berger (2009) claims that the association of   some SGRB with elliptical galaxies shows that their 
progenitors belong to an 
old stellar population, unlike LGRB. However  also  SGRB hosts have found to be  star forming galaxies.
The Swift satellite (Geherls 2004) greatly improved the understanding of SGRB 
progenitors  because revealing their location. 
At present, the majority of SGRB events  appear at relatively low z, at  $\sim$ 0.4-0.5 on average (Berger 2014).
This is an  important issue because it  could be  related  to the nucleosynthesis evolution  of the host 
galaxy stars  (Contini 2017a).
The lack of an associated supernova  in SGRB and the heterogeneous sample of host galaxies
(e.g. Kann et al. 2011) is consistent with a compact binary merger origin
(Rosswog et al. 2003), such as neutron stars or black holes.
The progenitors were related  to an old star population  compatible with 
NS (neutron star) -NS or NS-BH (black hole) encounters (Berger 2009, Eichler et al 1989).
The host galaxy analysis shows that long GRB
are found in star-forming galaxies (Fruchter et al 2006, Savaglio et al 2009 etc)
 while SGRB occur in both star-forming and early-type galaxies
(D'Avanzo et al 2015, Bloom et al 2002, Berger et al 2005, Fox et al 2005, Savaglio et al 2009, Fong et al 2013).
 SFR  in SGRB hosts are low and  a broad range 
of stellar masses is found in agreement with an old progenitor  adapted to the merging neutron stars
(Perley et al 2012).
The discovery of GW170817 the neutron star merger directly measured in gravitational waves 
and associated with a SGRB  allows to study  neutron star  mergers in general
 and to investigate, in particular, 
 whether the  gas within the NGC4993 host galaxy   has  peculiar characteristics 
 (Levan et al 2017,  Villar et al 2017, etc).

In previous papers  (e.g. Contini 2016) we have been investigating the physical conditions and the 
element abundances 
of GRB host galaxies by the analysis of the observed line  spectra at relatively high z.
We have compared them with those of various supernova (SN) types, active galactic nuclei (AGN), starbursts (SB),
HII regions, etc.
 In this paper we   investigate  line ratios and continuum SED of SGRB host galaxies. 
The modelling of the continuum  is 
less constraining than that of the line spectra, so   we chose  in our sample SGRB hosts that  have reported 
emission line fluxes.
 The sample is rather poor in number of objects.  The line spectra  suitable
to constrain  the  models generally contain at least [OII], [OIII], \Hb, \Ha~ and [NII].
The  \Ha/\Hb~  line ratios which are  used  to reddening correct the spectra cannot be neglected.

 In our model we   assume  that the continuum radiation, in terms of 
thermal Bremsstrahlung, originates from the same clouds of gas emitting the
line  fluxes.  
  Besides the effect of merging that affects both high redshift galaxies as well as local ones,
shock  waves are created by winds, jets and collisions throughout the galaxies. Therefore shocks
 should be accounted for by the  calculation of line and continuum fluxes.
In particular, shocks with velocities $>300$ \kms can heat the gas to high temperatures $>$ 10$^6$K 
which decrease following the cooling rate downstream. Consequently,  the  calculated
 Bremsstrahlung  covers the entire frequency range from radio to X-rays. 
 The flux from  the background old star population  emerges from the Bremsstrahlung SED  in the IR range 
in  local and high z galaxies (Contini 2018). Therefore,  
in order to reproduce the observed SED 
we  consider that  Bremsstrahlung  could  dominate  throughout
some significant frequency domains such as  the radio,  the optical-UV and the soft  X-ray, while
dust reprocessed radiation  appears in the far-IR.
Furthermore,  synchrotron radiation created by the Fermi mechanism at the shock front   
is often observed in the radio range.

 NGC4993 has enough  published  data  to allow line and continuum modelling.
For all  other SGBR hosts presented in this paper, the  continuum SED observations
cover only the IR range corresponding to emission from the  underlying stellar population. 
However,  for GRB050709, GRB100206A, GRB130603B  the
data for the optical-near IR line ratios reported by Berger (2009), Perley et al (2012), de Ugarte Postigo  
et al (2014),
Cucchiara et al (2013) and Soderberg et al (2006) are enough to constrain the models.
The code {\sc suma} calculates the continuum as well as the line fluxes. 
We will  show the continuum SED in the diagrams throughout a large frequency range (10$^8$-10$^{19}$Hz)
even if  we could  find  no data   in the radio, far-IR, optical-UV  and X-ray.
The Bremsstrahlung results are presented  as predictions.
For GRB111117A at z=2.21   no line fluxes are reported in the literature.
So  we will investigate only the old star  contribution  to the SED in order 
to compare it to   the SED of other SGRB hosts at lower z. 

The calculation code {\sc suma} is  presented in Sect. 2.  
 In Sect. 3 we report the detailed  modelling of line and continuum spectra  for each of the SGRB
host sample.
Concluding remarks follow in Sect. 4.

\section{Description of the calculations}

The code {\sc suma}  simulates the physical conditions of an emitting gaseous 
cloud under the coupled effect of photoionization from a radiation
source and shocks assuming a plane-parallel geometry  (Ferland et al 1995).
Two  cases  are considered relative to the cloud propagation : the photoionizing 
radiation reaches  the gas on the  cloud edge  corresponding to the shock front (infalling) 
or on the   edge  opposite to the shock front (ejection). 

To calculate the line flux and the continuum emitted from a gas the physical
conditions and the fractional abundances of the ions must be known.
In a shock dominated  regime
the calculations start at the shock front where the gas is compressed 
and thermalized adiabatically, reaching the maximum
temperature (T $\propto$ V$_s^2$, where \Vs is the shock velocity)
in the immediate post-shock region. Compression is 
calculated by  the Rankine-Hugoniot equations (Cox 1972) for the 
conservation of mass, momentum and energy throughout the shock front and downstream. 
Compression strongly affects the cooling rate and consequently,
the distribution of the physical conditions downstream, as 
well as that of the element fractional abundances.   
The downstream region is automatically cut in many plane-parallel slabs 
(up to 300) with different geometrical widths   in order 
to account for the temperature gradient throughout the gas.  Thus, the change of the physical 
conditions downstream from one slab to the next is minimal. 
In each slab the fractional abundances of all the ions is calculated resolving
the  ionization equations which
account for the ionization mechanisms (photoionization by the primary and diffuse radiation and
collisional ionization) and recombination mechanisms (radiative,
dielectronic recombinations) as well as charge transfer effects.
The ionization equations are coupled to the energy equation (Cox 1972),
when collisional processes dominate, and to the thermal balance equation if
radiative processes dominate. This latter balances the heating
of the gas due to the primary  and diffuse radiations reaching
the slab  and the cooling due to recombinations and collisional
excitation of the ions followed by line emission and
thermal Bremsstrahlung. The coupled equations
are solved for each slab, providing the physical conditions necessary
to calculate  the slab optical depth  and  the  line and
continuum emissions. The slab contributions are integrated
throughout the nebula.
The calculations stop when the electron temperature is as low as 200 K, if the nebula is 
radiation-bounded or  at a given value of the nebula geometrical 
thickness, if it is matter-bounded. 
The uncertainties of the calculation results are evaluated as $<$10 percent.

The main input parameters are  those referring  to the shock, which are the
preshock density n$_0$, the shock velocity \Vs, the magnetic field \B0
(for  all  galaxy models \B0=10$^{-4}$Gauss is adopted),
 as well as those characterizing 
the source ionizing radiation spectrum, and the chemical abundances 
of He, C, N, O, Ne, Mg, Si, S, Ar, Cl and Fe, relative to H. 
Generally, \Vs is constrained by the FWHM of the line profiles,
\n0 by the ratio of the characteristic lines.
The relative  abundances of the elements  are constrained by the line ratios.
In the case where shock and 
photoionization act on opposite sides of a plan-parallel nebula, 
the geometrical width of the nebula, $D$, is an input parameter.
The diffuse radiation bridges the two sides, and  the smaller $D$
 the more entangled are the photoionized and 
the shocked regions on the opposite sides of the nebula. In this
case, a few iterations are necessary to  obtain the 
physical conditions downstream.
The effect of dust present in the gas, characterized by the dust-to-gas
 ratio $d/g$  and the initial grain radius \agr are also  consistently
taken into account. 

The main characteristics of {\sc suma} are explained in detail in the following
sections.

\subsection{Photoionizing radiation flux}

The  radiation from a  photoionizing source is characterized by its
spectrum, which is calculated at 440 energies, from a few eV to KeV,
depending on the object studied. Due to  radiative transfer, the
radiation spectrum changes throughout the downstream slabs, each of them
contributing to the optical depth. The calculations assume a steady
state  downstream. In addition to the radiation from the primary
source, the effect of the diffuse secondary radiation created by the gas emission
(line and continuum) is also taken into account (see, for instance,
Williams 1967), using about 240 energies to calculate the spectrum.
The secondary diffuse radiation is emitted from the slabs of
gas heated  by the radiation flux reaching the gas and by the shock.
Primary and secondary radiation are calculated by radiation transfer.

 For an AGN, the primary radiation is the power-law radiation
flux  from the active centre $F$  in number of photons cm$^{-2}$ s$^{-1}$ eV$^{-1}$ at the Lyman limit
and  spectral indices  $\alpha_{UV}$=-1.5 and $\alpha_X$=-0.7. 
 $F$  is combined with the ionization parameter $U$ by
$U$= ($F$/(nc($\alpha$ -1)) (($E_H)^{-\alpha +1}$ - ($E_C)^{-\alpha +1}$)
(Contini \& Aldrovandi, 1983), where
$E_H$ is H ionization potential  and $E_C$ is the high energy cutoff,
$n$ the density, $\alpha$ the spectral index, and c the speed of light.

If the stars are the photoionization source
the number of ionizing photons cm$^{-2}$ s$^{-1}$ produced by the hot 
source is $N$= $\int_{\nu_0}$ $B_{\nu}$/h$\nu$ d$\nu$, 
where $\nu_0$ = 3.29$\times$10$^{15}$ s$^{-1}$ and B$_{\nu}$  is the Planck function. 
The flux from the star is combined with $U$ and n by $N$ (r/R)$^2$=$U$nc, where r is 
the radius of the hot source (the stars),
 R is the radius of the nebula (in terms of the distance from the stars), n is the density of the nebula and c is the 
speed of light. Therefore, \Ts  and $U$ compensate each other, but only in a qualitative way, because \Ts  determines 
the frequency distribution of the primary flux, while $U$ represents the number of photons per number of electrons 
reaching the nebula. The choice of \Ts and $U$   is obtained  by the fit of the line ratios.

In the turbulent regime created throughout a SB shocks are ubiquitous. 
Radiation from the stars photoionizes the gas.
For SBs we  assume for the primary radiation  a  black-body (bb) corresponding to an
 effective temperature \Ts and  a ionization parameter $U$.
 A pure bb radiation  accounting for \Ts is a poor approximation for a SB, 
even adopting a dominant spectral type (see Rigby \& Rieke 2004). 
Following Rigby \& Rieke,  "the starburst enriches and heats its ISM as well as the intergalactic medium.
The ionizing spectrum is set by the SB age, IMF and star formation history."  Adopting a single effective
temperature the entire SB field is represented by a single star type. However, 
the observed line spectra for high redshift galaxies at present  cover a  narrow  optical-near-IR range of frequencies,
the lines are few and from few ionization levels, therefore  the bb radiation flux calculated by a dominant
temperature is acceptable, also in view that
 the line ratios (that are  related  to \Ts) in a shock dominated regime also depend  on the
electron temperature, density, ionization parameter, metallicity, on the morphology of the ionized clouds, 
and in particular, they depend on the hydrodynamical field.
Therefore we  will determine  \Ts phenomenologically by  selecting the effective temperature \Ts which  leads 
to  the best fit of all the observed line ratios for each spectrum and we will use it to calculate the  continuum.

\subsection{Electron temperatures}

The temperature in each slab depends on energy gains (G) and losses (L) of the gas.
Close to the shock front downstream, collisional mechanisms prevail
and the temperature is calculated from the energy equation in terms of
the enthalpy change (Cox 1972).
In the slabs  where the temperature is $\leq$ 2 $\times$10$^4$ K, photoionization
and heating by both the primary and the secondary radiation dominate
and the temperature is calculated by thermal balance (G=L).
Gains are calculated by the rate at which energy is given to the electrons
by the radiation field (Osterbrock 1974).  The energy of suprathermal
electrons created by photoionization is rapidly distributed among the thermal
electrons through collisions, heating the gas.

Several processes contribute to the gas cooling. The cooling rate is:
L = L$_{ff}$+L$_{fb}$+L$_{lines}$+L$_{dust}$,
where L$_{ff}$ corresponds to Bremsstrahlung, particularly strong
at high temperatures and high frequencies.
Self-absorption is included in the calculations.
 L$_{fb}$ corresponds to free-bound losses due to recombination and
is high at T$\leq$ 10$^5$ K.
L$_{lines}$ is due to line emission  with the bulge between $\leq$ 10$^4$ K and 10$^5$ K.
 L$_{dust}$  represents the energy lost by the gas in  the collisional heating of dust
grains. It is high the higher $d/g$ and \agr.

Immediately behind the shock front the gas is thermalized to
a temperature of T=1.5 $\times$10$^5$(\Vs/(100 \kms))$^2$ K.
At high temperatures ($\geq$ 10$^6$ K) recombination coefficients are very
low.  The cooling rate is then low.
Considering that the cooling rate is $\propto$ $n^2$ (where $n$ is the density of the gas),
the thickness of the first slab depends strongly on $n$.

At T between 10$^4$ K and 10$^5$ K the UV lines and the coronal lines
in the IR are strong and lead to rapid cooling and compression
of the gas. If the cooling rate is so high to
drastically reduce the  temperature  eluding intermediate ionization-level lines,
the calculated spectrum will be wrong.
Therefore, the slab thickness must be reduced and all the physical
quantities recalculated. The choice of the slab thickness is determined
by the gradient of the temperature.
This process is iterated until the
thickness of the slab is such as to lead to an acceptable gradient of
the temperature (T(i-1) - T(i))/T(i-1) $\leq$ 0.1, 
where T(i) is the temperature  of slab i.)

As the temperature drops, a large region of gas with 
temperature  $\sim$10$^4$ K,  which is sustained by  the secondary radiation,
 is present in the radiation-bounded case,
i.e. when the gas recombines completely before reaching the 
 edge of the nebula opposite to the shock front. Due to a
lower temperature gradient,  calculations in this zone  
may be performed in slabs with a larger 
geometrical thickness. 

\subsection{Element abundances}

The absolute line fluxes  for the ionization level i of element K are calculated by the 
term $n_K$(i) which represents the density of the ion i. We consider that $n_K$(i)=X(i)[K/H]$n_H$, where X(i) is 
the fractional abundance of the ion i calculated by the ionization equations, [K/H] is the relative 
abundance of the element K to H and $n_H$ is the density of H (by number \cm3). In models including shock, 
$n_H$ is calculated by the compression equation (Cox 1972) in each slab downstream. 
So the abundances of the elements are given relative to H as input parameters.

\subsection{Dust reprocessed radiation}

Dust grains are coupled to the gas across the shock front by the magnetic field (Viegas \& Contini 1994). 
They are heated by radiation and collisionally by the gas to a maximum temperature which is a 
function of the shock velocity, of the chemical composition and of the radius of the grains, 
up to the evaporation temperature ($T_{\rm dust}$ $\geq$ 1500 K). 
The grain radius distribution downstream is determined 
by sputtering, which depends on the shock velocity and on the density. 
Throughout shock fronts and downstream, 
the grains might be destroyed by sputtering.
The grains are heated  by the primary and secondary radiation, and by gas collisional 
processes.  Details of the calculations of the dust 
temperature are given by Viegas \& Contini (1994). 
When the dust-to-gas ratio $d/g$ is high, 
the mutual heating of dust and gas may accelerate the 
cooling rate of the gas, changing 
the line and continuum spectra emitted from  the gas.
The intensity of dust reprocessed radiation in the IR  depends on $d/g$ and on the radius \agr.
In this work we use $d/g$=10$^{-14}$ by number for all the models which corresponds to
4.1 $\times$10$^{-4}$ by mass for silicates (Draine \& Lee 1994).

\begin{table*}
\centering
\caption{Modelling  NGC4993  observed line ratios [OIII]/\Hb=1.4 and [NII]/\Ha=1.26}  
\begin{tabular}{lcccccccccccc} \hline \hline
\ mod    &  [OIII]/\Hb     & [NII]/\Ha & \Hb    & \Vs  & \n0  & $D$ & \Ts       &  $U$ & log$F$ &12+log(O/H)&12+log(N/H)\\ 
\  -     & -               & -         &  $^1$     & \kms & \cm3 & pc  & 10$^4$K   &-     & $^2$      &-    & - \\ \hline
\ modSB  &  1.41           & 0.13      &37.7       & 300  & 300  & 1.67&  5        & 1    &  -     & 8.82   & 8.0\\ 
\ modSD  &  1.4            &0.146      &0.0009     & 100  &100   & 1.9 &  -        & -    &  -     & 8.82   &8.0\\
\ modAGN1&  1.43           &1.26       &0.05       & 100  &300   & 1   &  -        & -    &  9.55  & 8.82   &8.0\\
\ modAGN2&  1.45           &0.11       &16.9       & 300  & 300  & 0.27&   -       & -    & 9.48   & 8.82   &8.0\\  \hline
\end{tabular}

$^1$ in \erg calculated at the nebula;
$^2$ in photons cm$^{-1}$ s$^{-1}$ eV$^{-1}$ at the Lyman limit 
\end{table*}

\begin{figure*}
\includegraphics[width=8.6cm]{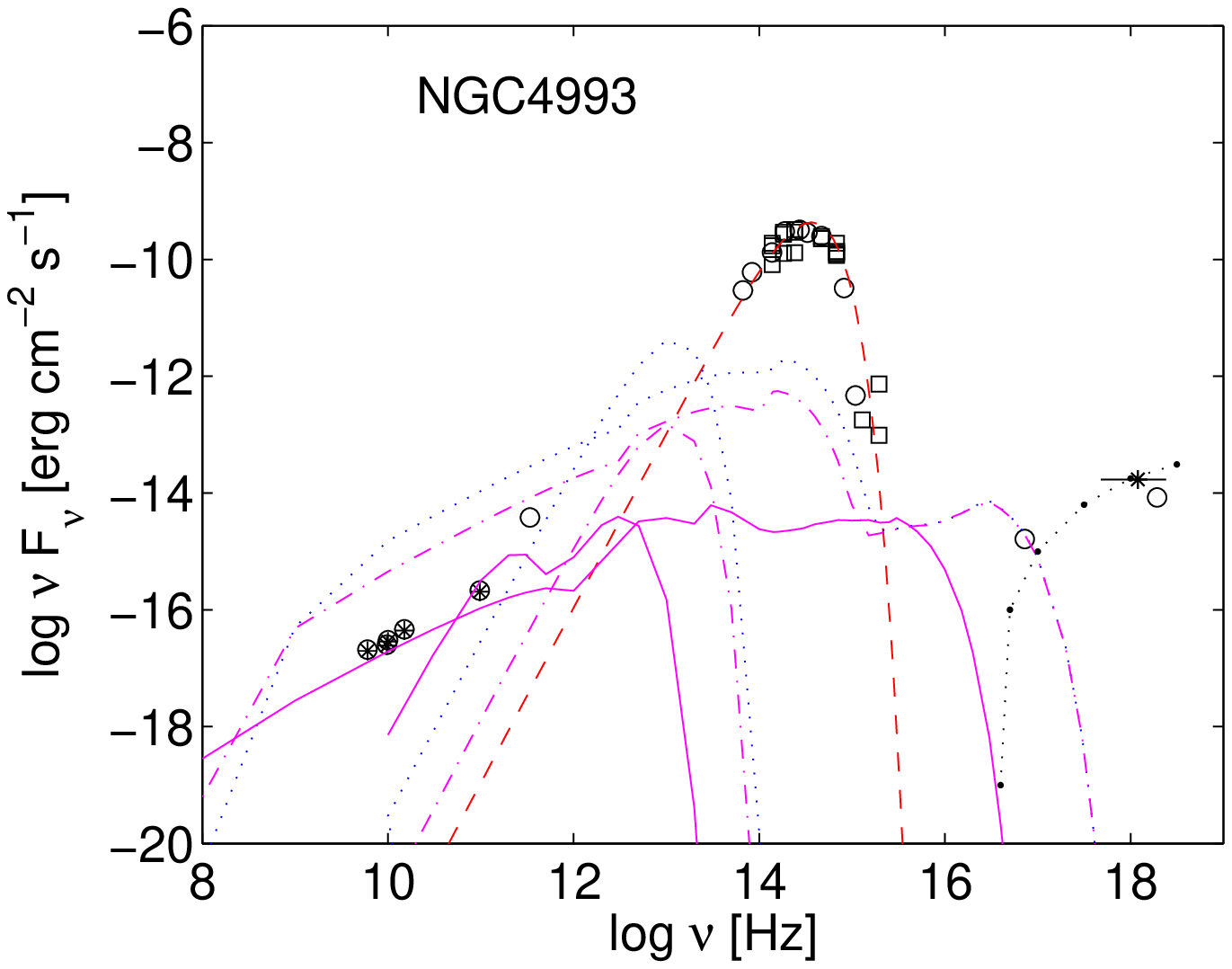}
\includegraphics[width=8.6cm]{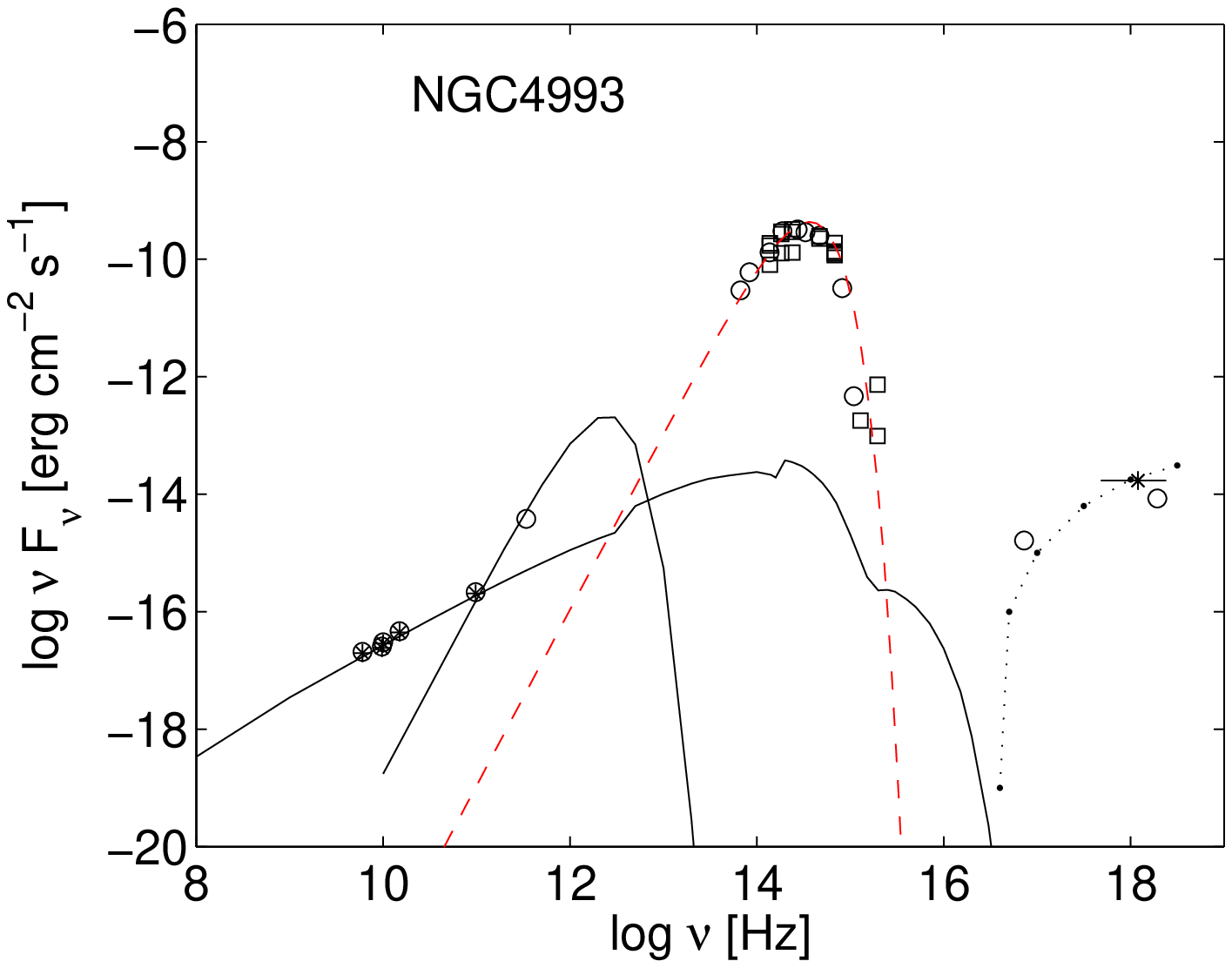}
 \caption{NGC4993  SED data : Wu et al (open circles); Levan et al (open squares); Blanchard et al IR-optical (asterisks):
 AGN (from NGC5252) flux  in the X-rays (dots).
 bb flux corresponding to 4000 K (red dashed).
Left : models modAGN2 (magenta dot-dashed), modSB (blue dotted) and modSD (magenta solid)
Right : model modAGN1 (black solid line) ;
other symbols as in Fig. 1 left diagram.  The models are described in Table 1.
}
\end{figure*}

\section{Modelling  host galaxy spectra}

The spectra observed from SGRB hosts are often poor in number of lines. 
BTP diagrams (Kauffmann 2003,  Kewley et al 2001) 
for the [OIII]5007/\Hb~ and [NII]6583/\Ha~ line ratios are  generally adopted
by the  author community in order  to   identify the galaxy  type in terms of the 
radiation source. 
However, extreme physical conditions and relative abundances far from  solar, in particular  for O/H and N/H, 
may  shift the observed [OIII]/\Hb~ and [NII]/\Ha~ line ratios throughout the BTP diagram
towards sectors which were  assigned to   different galaxy types.
For example, a low N/H relative abundance may shift the observed [NII]/\Ha~  line ratio emitted 
from a galaxy towards the SB galaxy domain  
although other features of the same object are characteristic of an  AGN.
This  may occur to GRB host galaxies which are  located at redshifts higher than local.
At high z, over- and under-abundances of the heavy elements are an important issue linked with 
merging, ages  and evolution.
So, to constrain a model when the line ratios are few
 we  follow a different  method.  [OIII]5007/\Hb~ and [NII]6548/\Ha~ data
are generally  observed because \Hb, \Ha, [OIII], [OII] and [NII] are  the strongest lines in the optical range.
A first hint  to the choice of the model is obtained by comparing the observed line ratios with the
grids of composite models (photoionization+shock) previously calculated (e.g. Contini \& Viegas 
2001a,b, hereafter CV01a and CV01b).
These models  provide an approximated but  rich information of the gas physical conditions calculated 
within the emitting clouds. (The line and continuum radiation flux are calculated at the nebula).
The line ratios  presented by the grids are adapted to   HII regions and AGN, respectively. 
The models (CV01a) corresponding to a black body  photoionization flux  can be applied 
to  SB galaxies (see Sect. 2). 
The [OIII]/\Hb~ and [NII]/\Ha~ line ratios alone cannot  definitively constrain the model because we deal 
with two line ratios 
related to two different elements. [OIII]/\Hb~ ratios depend mostly on  the ionization parameter and on \Vs 
but  the [NII]/\Ha~ ratios depend strongly on the N/H relative
abundances. For e.g. SN and GRB hosts N/H ranges between $\sim$ 0.1 solar and $\sim$5 solar (Contini 2017a).
We cannot determine a priori whether the best fit  to the observed [NII]/\Ha~ line ratio could be  reached by changing one or more 
input parameters
representing  the physical conditions (\Vs, \n0, $D$, $F$ for AGN, \Ts and $U$ for SB) or by modifying the 
N/H relative abundance. 
  N$^+$ and H$^+$ ions as well as  O$^+$ and H$^+$   are correlated by charge exchange reactions, therefore they
 have a similar trend throughout a cloud.
 When  [OIII]/\Hb~ (and [OII]/\Hb)  are well reproduced by modelling  and solar N/H are adopted, 
the resulting N/H is calculated by
([NII]/\Hb)$_{obs}$ = ([NII]/\Hb)$_{calc}$ (N/H)/(N/H)$_{\odot}$,  
where (N/H)$_{\odot}$ is the solar N/H relative abundance.
[OII]/\Hb~ and [OIII]/\Hb~ are more affected by the physical conditions of the emitting gas than by  O/H.

 On this basis we calculate a grid of models.  A  set of  models   (e.g. Table 1)
which best reproduce the line ratios is selected. We  obtain the final model by 
comparing the calculated SED with the observed one.

The calculated continuum SED  is represented in the diagrams by
 two  lines, one   for  the Bremsstrahlung which covers frequencies from radio to X-ray, and one
 for  the reprocessed radiation from dust in the IR range. They are calculated by the same model
which reproduces the line ratios. In the IR-optical range the background star flux emerges  from 
Bremsstrahlung and can be blended with dust reradiation when shock velocities are high.  
The errors in the observed photometric data are  $<$ 20 percent while the uncertainties in the 
calculations are $<$ 10 percent.

 In Sect. 3.1 we reproduce by detailed modelling the NGC4993 observed  line ratios and the continuum SED
(Levan et al 2017, Palmese et al. 2017, Wu et al 2017).
In Sect. 3.2  we present  modelling results of the line spectra observed from the host galaxies
included in the  Berger (2009) SGRB sample.
The calculated continuum  is  compared  to the SED  observed by
Savaglio et al (2009) in the IR for SGRB050709.
In Sect. 3.3 we  show the modelling of  line and continuum
spectra  reported  by  Perley et al  (2012) for SGRB100206A host.
In Sect. 3.4 the line ratios observed from the
SGRB130603B (de Ugarte Postigo et al 2014) and SGRB051221a (Soderberg et al 2006) host galaxies  and   modelled
by Contini (2016) are reported. SGRB130603B has been observed  in the IR  allowing the  modelling of the SED.
 In Sect. 3.5 the  modelling of the  SGRB111117A continuum  observed by Selsing  et al (2017)  in the IR
 is shown.

\subsection{NGC4993 : the host galaxy of GW170817}

Observations by Ligo and Virgo reported  gravitational waves which followed to the  recent  
neutron star merger  event  GW170817.
In the frame of our  analysis of GRB host line and continuum SED we calculate the characteristics of the 
emitting gas from NGC4993 which is considered to be   the host galaxy of  GW170817 at z=0.009873.

Levan et al (2017)  in their fig. 4  compared   [OIII]/\Hb~ and [NII/\Ha~  
spectroscopic observations with  BTP diagnostics.
They   found that  the line ratios fitted  the AGN domain.
Spectroscopic observations  by Palmese et al (2017, fig. 3 right panel) show [OIII]5007/\Hb~  and
[OII]/\Ha~ within the error of Levan et al (2017) observations. 
However, Palmese et al  admitted that the [OIII]/\Hb~ line ratio is very uncertain.
In Table 1 we  report the calculated  models which best reproduce  Levan et al  spectroscopic data.
For all of them O/H solar (6.6$\times$10$^{-4}$) and N/H solar (10$^{-4}$) are adopted (Grevesse \& Sauval 1998).
Model
modSB is characterised by a bb photoionizing radiation (suitable to  SB and HII regions)
 and relatively high shock velocities (\Vs=300 \kms), 
modSD is a shock dominated model ($F$=0 and $U$ =0) with \Vs=100\kms. Models
modAGN1 and modAGN2 are  calculated by a power-law photoionizing flux and different shock velocities
  \Vs=100 \kms and \Vs=300 \kms, respectively.

 Wu et al (2017) present new Very Large Array (VLA) and ALMA data and use them for broad band modelling.
The continuum SED  is shown in Fig. 1. Haggard et al (2017)  report
Chandra observations which reveal a compact source consistent with the nucleus of the galaxy and L$_X$
$\sim$ 2$\times$10$^{39}$ erg s$^{-1}$ (0.5-8 keV).
Wu et al suggest that the X-ray soft energy excess may indicate  thermal emission from a gaseous component in the galaxy.
 Some observations in the IR were also available
(Lambert \& Valentijn 1989, DeVaucouleurs et al 1991, 2MASS 2003). They are  used to constrain the model.

 In  Figs. 1-4 accounting for the continuum SED, each model corresponds
to two  curves (Bremsstrahlung and dust reradiation). The same symbol is used for  both.
The models  which reproduce the SED in Fig. 1 are  those used to fit the line ratios in Table 1.
Shock velocities \Vs= 300 \kms which correspond to a maximum temperature downstream T $\sim$ 10$^6$K  
shift the  Bremsstrahlung  maximum to relatively high frequencies and fit the datum at 7.25 $\times$10$^{17}$ Hz  
(Fig. 1, left diagram), 
corresponding to  soft X-ray.   However, the continuum SED in the radio-FIR range calculated by   models
modSB and modAGN2 overpredicts the data in the radio range
by a factor of $\sim$ 10.  So we dropped modSB and modAGN2.
The shock dominated model (modSD) calculated  with \Vs=100 \kms and \n0=100 \cm3 could fit the radio-FIR
data  adopting a low  $d/g$ = 4 $\times$ 10$^{-5}$. In this case the data in the X-ray have no valid explanation.
Moreover, the [NII]/\Ha~ ratio is  underpredicted by a factor of 10. To adopt  N/H relative abundance 10 times solar
in order to fit the [NII]/\Ha~ ratio is rather an extreme solution. Therefore we select   modAGN1 as the best 
reproducing  model regarding  both the line ratios and the SED  of  NGC4993.
The model  is calculated with \Vs=100 \kms and \n0=300 \cm3. These shock velocities and densities are  found
in the narrow line region (NLR) of AGN.  The power-law flux is relatively low,  suitable to low-luminosity AGN 
(LLAGN, e.g. Contini 2004) and LINERs.

As already mentioned we calculate the continuum at the nebula whereas the data are observed at Earth.
The same is valid for the \Hb~ line flux.
To   reduce the calculated  \Hb~ flux  by the distance  galaxy - Earth, we combine
  the \Hb~  flux observed at Earth  with  \Hb~  calculated at the nebula by :

 \Hb$_{calc}$$\times$r$^2$$\times$ \ff= \Hb$_{obs}$$\times$d$^2$.
 \Hb$_{calc}$ is given in Table 1.  \Hb$_{obs}$  is    obtained from the Kennicutt (1998) relation
linking SFR with \Ha,  adopting  \Ha/\Hb=3 and  SFR = 0.003 \msol y$^{-1}$  (Pan et al 2017).  
\ff~ is the filling factor,
r  the distance of the nebula from the  {radiation source} and d the distance of the galaxy to Earth.
We obtain  the factor \fs=(r/d)$^2$ = \Hb$_{obs}$ /\Hb$_{calc}$/\ff~  which  we use in Fig. 1 to  shift
the calculated continuum SED in order  to be  compared  with the data observed at Earth. In fact
 we consider that the Bremsstrahlung is emitted from the same cloud which emits the lines (e.g. \Hb). 
From Fig. 1 (right diagram) the best fit in the radio-NIR range is obtained by log(\fs)= -11.7,
while log(\Hb$_{obs}$/\Hb$_{calc}$)= -14.  
Then we obtain \ff=0.005.   
The distance from Earth d=42.9 Mpc, then r results 60 pc.
To reproduce the  dust reradiation bump in Fig. 1 (right) we  
adopted $d/g$=0.008 and \agr=0.1 \mum.

Considering that an AGN is present in the NGC4993 galaxy, we try to fit the X-ray data presented by Wu et al
by the (absorbed) power-law model which was successfully adopted to  reproduce the X-ray at high frequencies 
observed for other AGN e.g.  the Circinus galaxy (Contini, Prieto \& Viegas 1998a), 
NGC5252 (Contini, Prieto \& Viegas 1998b), etc.
The observed X-ray flux from the NGC4993 AGN nucleus  is relatively low, exceeding
 the maximum radio  flux  by a factor of $\sim$100 (Fig. 1 right), while in e.g. the Circinus AGN 
this factor is $>$ 10$^5$ and in NGC5252 $\geq$1000.
Wu et al claim that this X-ray emission is mostly due to a weak LLAGN.
The results agree with
Palmese et al (2017) who believe that the stellar model fit to NGC4993 reveals the existence of weak ionized gas emission
while the line ratios are most probably produced by a harder ionizing source than star formation because
lying in the AGN region of the Baldwin
et al (1981) diagram. Blanchard et al (2017) argue that there is a weak AGN present in the core of the galaxy
 on the basis of radio and X-ray emission in agreement  with  Wu et al (2017)  who suggest
 a LLAGN or even some sort of shocks.
 Palmese et al concluded that there is no evidence
of recent star-formation from the  spectrum, irrespective of the uncertain [OIII]/\Hb~.
From the Balmer decrement they found  E(B-V)=0.12$\pm$0.50. Dust obscuration does not play a role in SFR estimation
 calculated on the basis of the \Ha~ luminosity.

Summarising, the
 observed SED presented by   Wu et al  shows soft and hard X-ray emission at log($\nu$[Herz]) $\geq$ 17.8.
Our  modelling confirms that velocities throughout NGC4993 are $\sim$ 100 \kms.
Therefore, both soft and  hard X-rays  (Fig. 1, left diagram) come from the AGN  radiation.
On the other hand, comparison with the   SED  calculated with \Vs $\geq$300 \kms
   could  nicely fit  NGC4993 data in the
soft X-rays domain. But  this model is dismissed because overpredicting the data in the radio range.
 Moreover,  it corresponds to [NII]/\Ha=0.16.
 Fig. 1  shows that  bb radiation flux from the background population stars corresponding to \Tbb=4000K  is  
  high relative to  Bremsstrahlung.

Concluding with Palmese et al., NGC4993 experienced a  minor galaxy merger with still visible signs.
The  spectral analysis shows AGN activity and an old stellar population.
In the IR-optical domain the bb radiation corresponding to the old star background population at 4000 K
reproduces  the data satisfactorily. 
 Compared to local SB and AGN galaxies (Contini \& Contini 2007), the bb flux-to-Bremsstrahlung ratio 
of $>$ 100 for old stars in the IR-VIS frequency range is remarkably high.

\subsection{Berger (2009) sample}

The Berger sample of SGRB hosts  is relatively  poor in number of objects.
In Table 2 we present the observed SGRB host galaxy line fluxes. 
The observed line   ratios to \Hb~ (in the underlying row) for each object are followed in the next 
row by the calculated ones.
 The models, represented by the set of the main input parameters which leads to the best fit, 
are shown in Table 3.
In the last column of Table 3 the metallicities evaluated by Berger (2009) are shown.
The lines are observed at Earth but calculated at the emitting nebula.
  We  compare the calculated line ratios  to the  observed line ratios to \Hb~ corrected for reddening.
For SGRB061006 and SGRB070724  \Ha~ and [NII] are lacking  and the line ratios to \Hb~ are modelled directly 
without reddening correction.   The  [OII] 3727+  doublet which is close to  the near UV domain  is  more affected
by correction than the other observed lines in the optical-IR. 
 Nevertheless, the ensemble of the parameters which  represents the model should be taken  with caution.
GRB51221a will  be shown in Sect. 3.4.
Berger (2009) analysing the spectra by the strong line methods, found that short GRB host 
luminosities are systematically brighter than  those of long GRB in the same redshift range.
Berger  obtained luminosities L$_B$$\sim$0.1-1.5 \Lsol , SFR $\sim$0.2-6 \msol yr$^{-1}$ and metallicities  
12+log(O/H) $\sim$8.5-8.9 (3.2$\times$10$^{-4}$ and 7.9$\times$10$^{-4}$, respectively)
higher than for LGRBs.
Specific SFR for SGRB have a median value lower than for LGRB SFR hosts.
Comparing the results obtained for the Niino et al (2016) sample of LGRB hosts  (Contini 2017b)  to the present ones 
in Tables 3 and 5, we obtain  similar results.  

 We have selected SGRB050709 and SGRB050724 photometric data from the Savaglio et al (2009) sample
to model the SED. They  show at least 3 datapoints.
However, for SGRB050724 we could not find  any of the observed line ratios  generally used to model the host galaxy.
 For SGRB050709 we  calculate the continuum SED by model modb4 (Table 3) at the nebula. 
  Savaglio et al data  for SGRB050709  are reproduced by a stellar background bb at \Tbb=3000 K in Fig. 2. 
Proceeding as for NGC4993 (Sect. 3.1)  we combined \Hb$_{calc}$  (Table 3) with
\Hb$_{obs}$  presented by Berger (2009, table 2). Adopting \ff=0.01 we calculated \fs and
 accordingly we shifted the  calculated continuum in Fig. 2.
Hopefully future data will confirm the modelling.
The distance  of the emitting nebula from the photoionizing source r results 144 pc.

\begin{table*}
\centering
\caption{Modelling Berger (2009) SBLG spectra}
\begin{tabular}{lcccccccccccccccc} \hline \hline
\ GRB        &z         &[OII] & \Hb    &  [OIII]   & \Ha  & [NII] & ref \\ 
\            &          &3727  & 4865   &  5007+    &6563  & 6548+ &     \\ \hline
\ 061006$^1$ &0.4377    & 4.1  & 0.9      &1.5      &  -   &  -    &   Berger (2009) \\
\ 061006$^2$ &          & 4.24 & 1        &1.89     &  -  &  -     &            \\
\ modb1      &          & 4.5  & 1        &1.86     &  -  &   -    &         \\
\ 061210$^1$ &0.4095     & 22.  & 7.6     &10.5     & 11.  & 2.4   &Berger (2009)\\
\ 061210$^3$ &           & 1.71 & 1       &1.6      & 3    & 0.15  &           \\
\ modb2      &           & 1.77 & 1       &1.57     & 2.98 & 0.15  &            \\
\ 070724$^1$ &0.4571     & 37.  & 15.     &17.      & -    &   -   &Berger (2009) \\
\ 070724$^2$ &           & 2.47 & 1       &1.13     & -    &   -   &         \\      
\ modb3      &           & 2.54 & 1       &1.18     & -    &   -   &           \\
\ 050709$^1$ &0.1606     &  -   & 6.6     & 26.     &26.   & 1.8     &Fox et al (2005)\\  
\ 050709$^3$ &           &  -   & 1       &3.7      &3.    & 0.29    &        \\
\ modb4      &           & (1.69)& 1      &3.78     & 3    & 0.29    &        \\ \hline

\end{tabular}

 $^1$ line flux in 10$^{-17}$\erg observed at Earth; $^2$ line ratios to \Hb ; $^3$ corrected line ratios to \Hb

\centering
\caption{Models for Berger (2009) SBLG spectra}
\begin{tabular}{lcccccccccccccccc} \hline \hline
\ mod       & \Vs    &\n0  & $D$         & 12+log(N/H)& 12+log(O/H) & \Ts      &  $U$ & \Hb    & 12+log(O/H)(Berger) \\ 
\           & \kms   &\cm3 & 10$^{18}$cm &           &              &  10$^4$K &  -   & $^1$   &     \\ \hline
\ modb1   &  100     & 100     & 0.8       & 7.       & 8.82     & 5.7      & 0.005   & 0.006 &8.63  \\
\ modb2   &  100     & 300     & 2.8      & 6.9      &  8.8     &  4.3       &0.03    &0.167  &8.82  \\  
\ modb3   &  120     & 250     & 3.       &   -      &  8.81    &  4.9       &0.013   &0.072  &8.88  \\
\ modb4   &  120     & 250     & 3.       & 7.25     &  8.81    &  4.8       &  0.06  &0.19   &8.50  \\ \hline

\end{tabular}

$^1$ in \erg calculated at the nebula; 

\end{table*}

\begin{table*}
\centering
\caption{Modelling Perley et al (2012) short GRB100206A host spectra at z=0.4068}
\begin{tabular}{lcccccccccccccccc} \hline \hline
\                 & [OII] &   \Hb   & [OIII]   & \Ha  & [NII]   & [SII]  &[SII] \\
\                 &3727   & 4865    &  5007+   &6563  & 6548+   & 6716   & 6731 \\ \hline
\ subtr$^1$           &19.6   &21.8     & 6.8      &112.8 & 58.8    & 20.4   & 13.6  \\
\ F/\Hb           & 0.9   &1        &0.31      &5.17  & 2.7     & 0.93   & 0.62  \\
\ corr            &1.34   & 1       &0.23      &3.    &1.58     & 0.53   &0.35   \\
\ modp1           &1.6    & 1       &0.23      &3.    &1.54     & 0.43   &0.35   \\ 
\ North (subtr)$^1$   & 7.91  & 9.59    &1.22      &33.41 &16.98    & 4.44   & 4.0   \\
\ F/\Hb           & 0.82  & 1       &0.127     &3.48  & 1.77    & 0.46   & 0.42   \\
\ modp2           & 0.82  & 1       &0.127     &3.12  & 1.6     & 0.4    & 0.42  \\ 
\ Center (subtr)$^1$  & 5.69  & 7.57    &3.5       &60.55 & 33.9    & 3.33   & 3.24  \\
\ F/\Hb           & 0.75  & 1       &0.46      &8.    & 4.47    & 0.44   &0.43   \\
\ corr            & 1.57  & 1       &0.27      &3.    & 1.67    & 0.15   &0.149  \\
\ modp3           & 1.64  & 1       &0.28      &3.2   & 1.4     & 0.14   &0.14   \\ 
\ South (subtr)$^1$   & 4.02  &3.61     &1.68      &17.7  &10.54    & 3.06   &4.94   \\
\ F/\Hb           & 1.11  &  1      &0.465     &4.9   &2.92     & 0.85   &1.37   \\
\ corr            & 1.6   & 1       &0.36      &3.    &1.8      &0.5     &0.81   \\
\ modp4           & 1.57  & 1       &0.367     &3.2   &1.8      &0.51    &0.61   \\ \hline

\end{tabular}

$^1$ in 10$^{-17}$ \erg

\centering
\caption{Models for Perley et al (2012) GRB100206A spectra}
\begin{tabular}{lcccccccccccccccc} \hline \hline
\       &  \Vs  &\n0  &   $D$      &12+log(N/H)&12+log(O/H)&12+log(S/H) &\Ts     &  $U$   &   \Hb       \\ 
\       &  \kms &\cm3 &10$^{18}$cm &           &           &            & 10$^4$K & 0.001  &  $^1$           \\ \hline
\ modp1 & 100   &120  &4.9         &8.04       &8.78       &6.48        &4.4      &0.9     &0.003             \\
\ modp2 & 100   &250  &3.3         &8.25       &8.82       &6.95        &3.6      &0.6     &0.025            \\
\ modp3 & 100   &200  &3.3         &8.04       &8.82       &6.40        &4.6      &0.6     &0.0098           \\
\ modp4 & 120   &200  &1.9         &8.15       &8.82       &7.11        &4.       &0.6     &0.0092           \\ \hline

\end{tabular}

$^1$ in \erg calculated at the nebula

\end{table*}

\begin{figure}
\includegraphics[width=8.6cm]{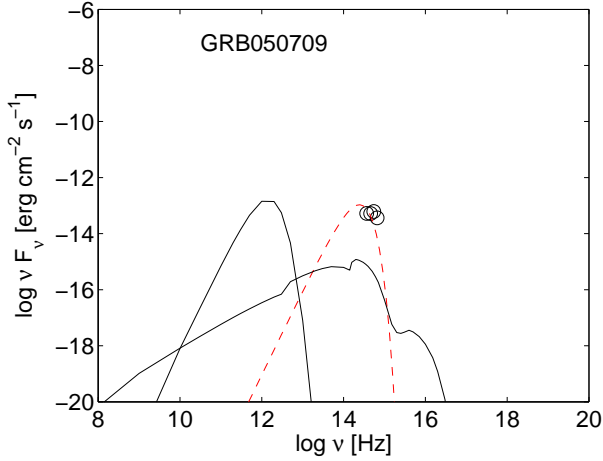}
 \caption{The SED of  SGRB050709 host from the Savaglio et al sample.
Open circles : the data;  black solid lines :   model modb4 (Table 3);
dashed line : bb flux corresponding to \Tbb=3000K.
}
\end{figure}

\begin{figure}
\includegraphics[width=8.8cm]{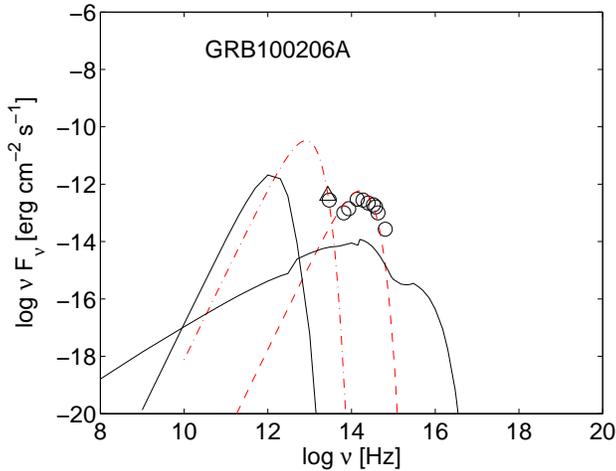}
\caption{The SED of SGRB100206A  host from the Perley et al sample. Open circles : the data;
black solid lines : model modp1 (Table 5);
dashed line :  bb flux corresponding to \Tbb=2000 K;
dash-dotted line  : bb flux corresponding to   \Tbb=100 K.
}
\end{figure}

\begin{figure}
\includegraphics[width=8.8cm]{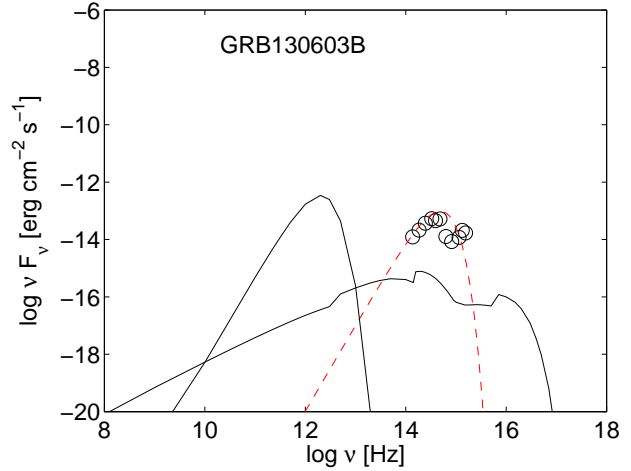}
\caption{The SED of GRB130603B host  from the de Ugarte Postigo et al  sample.
Open circles : the data;
black solid lines : model MS5 (Table 6);
dashed line : bb flux  corresponding to \Tbb=6000 K.}
\end{figure}

\begin{figure}
\includegraphics[width=8.6cm]{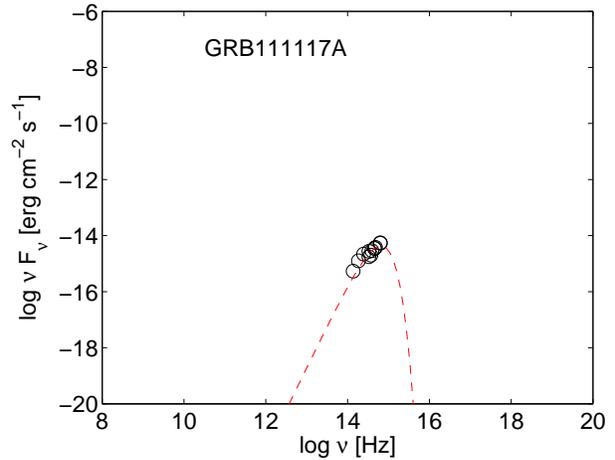}
\caption{The SED of GRB111117A host  from Selsing et al.
Open circles : the data;
dashed line : bb flux corresponding to \Tbb=8000 K.
}
\end{figure}

\subsection{Spectra of short GRB100206A  host by Perley et al (2012)}

Perley et al (2012) presented data from Swift for GRB1002016A (Table 4)   a disk galaxy  
at z=0.4068 rapidly forming stars.
The galaxy is red, obscured and the interpretation of the spectra  reported by 
Perley et al  leads to  a high metallicity (12+log(O/H)=9.2).
They also explain the SED by a substantial stellar mass of older stars although the  IR luminosity
(4$\times10^{11}$\Lsol) could indicate young star formation.
Perley et al  show a single spectrum and  the spectra observed  in different locations with the particular aim to
determine the metallicity (in term of O/H).
We present  in Table 5 the physical conditions  calculated by the detailed modelling 
of the spectra presented by Perley et al (2012) in their tables 2 and 3.
Unfortunately  the lines are few, nevertheless they are the minimum  required to constrain the models.
Our results show relatively high preshock densities  in the emitting clouds which are  more extended
than those in  LGRB hosts (Contini 2017a).  We find solar O/H 
(12+log(O/H)=8.82) in  nearly 
all the observed positions, while N/H are higher than solar by a small factor ($<$2), but higher 
by a factor of $\sim$ 10
than for the SGRB hosts in the  Berger sample. The N/O ratios are 
the highest  ever calculated  for  SGRB, even  higher than those calculated for  the de Ugarte 
Postigo et al (2014) GRB spectra and   those calculated for long GRB at the same redshift. 
The stellar background contribution to the SED corresponds to \Tbb=2000 K.
The calculated SED is shown in Fig. 3.

\subsection{SGRB130603B (de Ugarte Postigo 2014) and SGRB051221a (Soderberg et al 2006)}

\begin{table*}
\centering
\caption{Models for  SGRB 130603B  at z=0.356 and SGRB 051221a at z=0.546 host galaxy spectra}
\begin{tabular}{lcccccccccccccccc} \hline  \hline
\              &OT site$^1$&mS1    &OT site$^2$&mS2  &core$^2$&mS3&  arm$^2$ &mS4&obs$^3$ &mS5 & 051221a$^4$&mS6\\ \hline
\ [OII]3727+   & 4.47      & 4.5   &3.4        & 3.6 &3.55&3.7    &4.3   & 4.3   &3.05& 3.2    & 7.8  & 7.5\\
\ H${\gamma}$  &  -        &0.46   &0.89       & 0.49&0.49&0.46   &[0.5] & 0.46  & -  &0.46    & 0.46 & 0.46\\
\ \Hb          &  1        &1      &1          &1    & 1  & 1     &1     &1      &1   &1       & 1    &1 \\
\ [OIII[5007+  &  0.87     & 0.9   &0.76       & 0.8 &0.75&0.8    & 0.85 & 0.85  &0.77& 0.7    &5.17  &5.15 \\
\ \Ha          &  3.       & 3.    &3.         & 2.96&3.  &2.96   & 3.   & 3.    & 3. & 3.     &-     &3.  \\
\ [NII]6585    &  0.7      & 0.76  &0.57       & 0.56&0.85&0.85   & 1.8  & 1.8   &0.78& 0.8    &-     &-\\
\ [SII]6717    &  0.66     & 0.7   &1.19       & 1.1 &0.63&0.66   & 0.27 & 0.5   &-   & -      &-     &-\\
\ [SII]6731    &  0.33     & 0.6   &0.56       & 0.97&0.5 &0.59   &  0.5 & 0.45  &-   & -      &-     &-\\
\   \Vs (\kms) &   -       & 140   &   -       &150  &  - &150    & -    & 140   &-   & 120    &-     &150\\
\  \n0  (\cm3) &  -        & 120   &   -       &130  &  - &130    & -    & 120   &-   & 100    &-     & 100\\
\  \B0  (10$^{-4}$G)&  -   & 3     &   -       &3    &  - &3      & -    & 3     &-   & 2      &-     & 1\\
\ $D$  $^5$    &  -        & 5.3   &   -       &1    &  - & 1     & -    & 1     &-   & 3      &-     & 0.2\\
\ 12+log(O/H)  &  -        & 8.82  &   -       &8.8  &  - & 8.8   & -    & 8.82  &-   & 8.81   &-     & 8.82\\
\ 12+log(N/H)  &  -        & 7.3   &   -       &7.3  &  - &7.48   & -    & 7.8   &-   & 7.48   &-     &7.11 \\
\ 12+log(S/H)  &  -        & 6.6   &   -       &7.0  &  - &6.78   & -    & 6.48  &-   & 6.48   &-     &6.48\\
\ \Ts (10$^4$K)&  -        &3.6    &   -       &3.5  &  - &3.5    & -    & 3.6   &-   &3.8     &-     & 8.3\\
\ $U$          & -         &0.014  &   -       &0.01 & -  &0.01   & -    & 0.014 & -  &0.007   &-     &0.007\\
\ \Hb $^6$     &  -        & 0.014 &   -       &0.016&  - &0.016  & -   &  0.03  & -  & 0.014  &-     & 0.007\\ 
\ 12+log(O/H)$^7$ &  -        &7.94-8.98& -       &8.16-8.88&-&8.17-8.87&-  &8.26-8.77&- & 8.5    &-     & 8.7  \\ \hline
\end{tabular}

$^1$de Ugarte Postigo et al (2014)(X-shooter); $^2$de Ugarte Postigo et al (2014)(FORS);$^3$ Cucchiara et al (2013)(DEIMOS); 
$^4$ Soderberg et al (2006)(Gemini-N GMOS); $^5$ in 10$^{18}$ cm;  $^6$ in \erg; $^7$  evaluated by  the
observers adopting  the strong line method.

\end{table*}

 \begin{figure*}
\includegraphics[width=8.8cm]{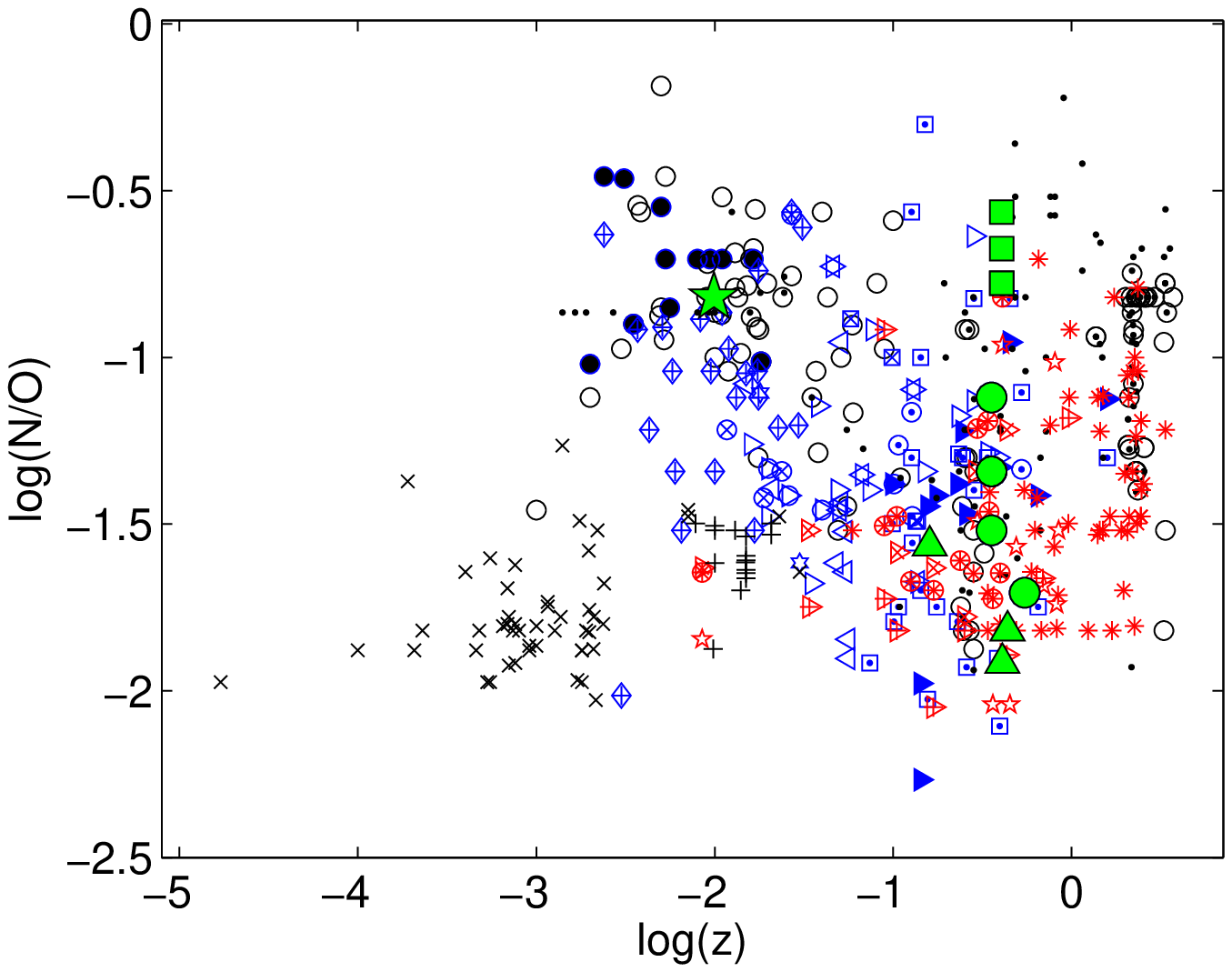}
\includegraphics[width=8.8cm]{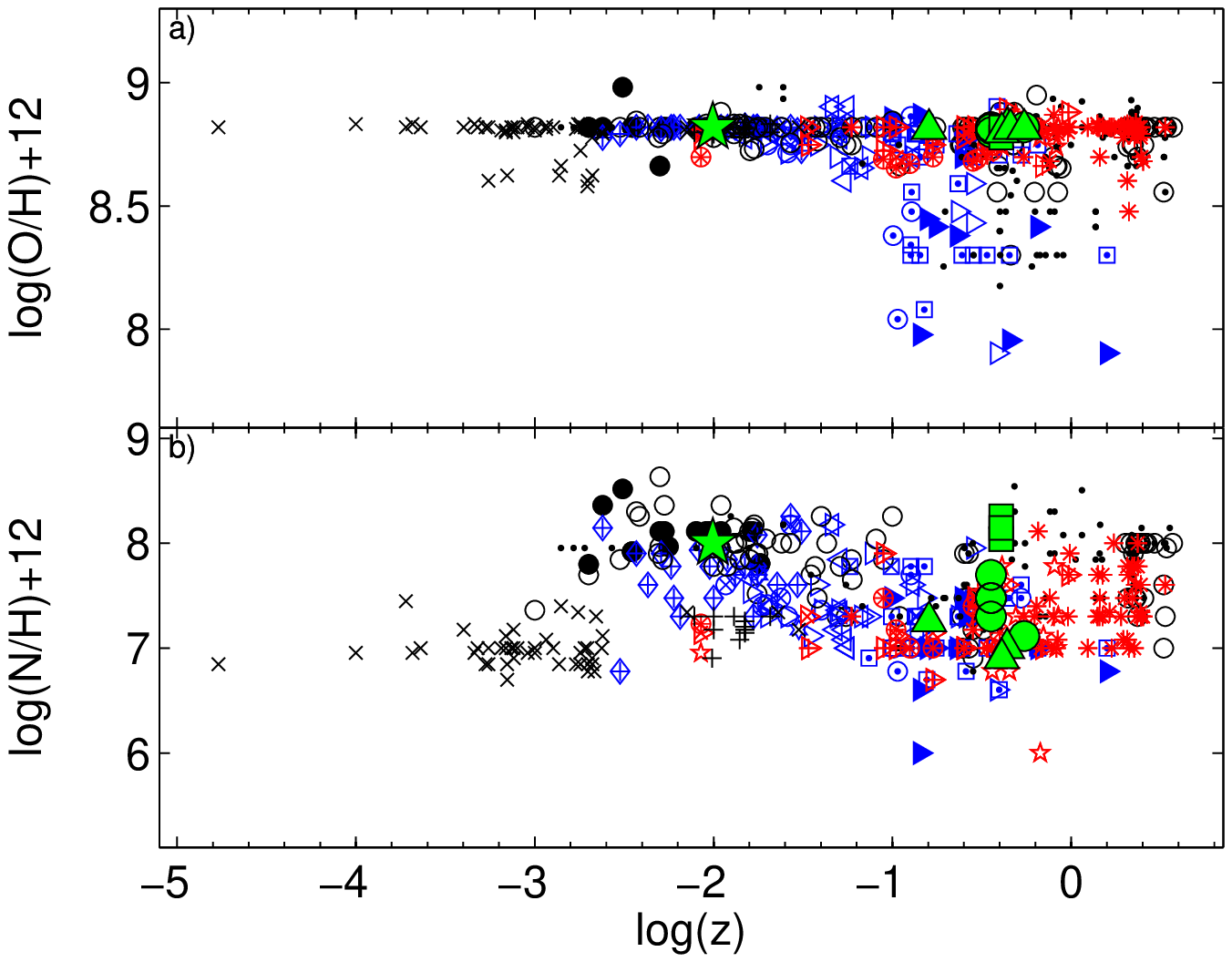}  
\caption{Left: distribution of N/O calculated by detailed modelling for SGRB hosts as 
function of the redshift 
(large circles : from the spectra observed by  de Ugarte Postigo et al; large squares : from
Perley et al; large triangles : from Berger;
large star: NGC4993). The results obtained in the present paper
are superposed  on those calculated for SN host galaxies (small triangles, filled diamonds,  points inside a  square),
long GRB hosts (asterisks), short GRB hosts (stars),  
SB (dots), AGN (open circles) and LINERs  (filled circles) both local and at higher z;
plus and cross : HII region galaxies nearby and local, respectively.  
Right : the same for O/H and N/H relative abundances.
 Symbols as in  the left diagram; 
References are given by  Contini (2017a, table 9).
}                                                                                      
\end{figure*}

 We report in the following the modelling of SGRB130603B and SGRB051221a line ratios
by Contini (2016). We present in Fig. 4 the  modelling of the SED. 
The spectra by FORS2 and X-shooter at VLT and ACAM at WHT were observed
by de Ugarte Postigo et al (2014). They   suggested   that SGRB derive from the merger 
of compact objects,
in particular  SGRB 130603B, on the basis of the detection of "kilonova"-like signature
associated with {Swift}. The host galaxy  
is a perturbed spiral due to interaction with another galaxy
(de Ugarte Postigo et al). In  the spectrum taken by X-shooter the afterglow dominates the continuum, but
 the  lines  are emitted from the host   therefore they are used  for modelling. 
 FORS spectra   show the core, the  arm and the opposite side of the galaxy.
 The GRB is located in the outskirts of the galaxy in a tidally disrupted arm at $\sim$ 5.4 kpc
from the brightest point of the host. 
The host was covered by two slit positions along the major axis and two spectra were taken at 
different times plus a X-shooter spectrum covering the afterglow position.
The results of modelling the X-shooter and FORS reddening corrected
spectra observed by de Ugarte Postigo et al 
(see Contini 2016, table 12) are reported in Table 6,  neglecting the GTC spectra because they do not 
include the \Ha~ line.
The spectrum observed by Cucchiara et al (2013) for SGRB 130603B is also  shown for comparison.
Most of the spectra presented  by SGRB surveys (e.g. Fong et al 2013) do not contain
enough lines to constrain the models.
To best reproduce all the line ratios presented in Table 6
 S/H lower than solar ((S/H)$_{\odot}$= 1.6$\times10^{-5}$, Grevesse \& Sauval 1998)
  by  factors $\geq$ 2 were adopted. S/H in
FORS (OT site) is near solar.
 The cloud geometrical thickness  results relatively large  ($D$=1.8 pc) from the  modelling of
the X-shooter (OT site) results. $D$=0.3 pc  is used to fit the  FORS spectra, showing a larger cloud fragmentation.
The spectrum reported by Cucchiara et al (2013) is reproduced by
  model (mS5) similar to those used to fit the spectra  observed  by de Ugarte Postigo et al.
The SB temperatures and ionization parameters  which result from  modelling  the  host spectra
are  relatively low 
(\Ts$\sim$ 3.5$\times$10$^4$ K and $U$$\sim$ 0.01).

The spectrum observed by Soderberg et al (2006) for SGRB051221a at z=0.546
and  the relative model mS6 are reported  from Contini (2016, table 12, last two columns) in Table 6.
The line ratios were reddening corrected adopting H${\gamma}$/\Hb=0.46.
The best fitting model shows 12+log(O/H)=8.8, in agreement with Soderberg et al who  obtained 8.7  
by the R32 method (upper branch). We have found \Ts=8.2$\times$ 10$^4$K and $U$=0.007.
\Ts  is higher than for SGRB130603B, where  \Ts  $\sim$ 3.5$\times$ 10$^4$ K. $U$, however,  is similar.
 Low ionization parameters (i.e. the number of photons reaching the nebula per number of electrons at the nebula)
roughly indicate that the  host observed positions  are far from the star-forming region
or that the photon flux is prevented from reaching the nebula.
The  hosts  show N/H $\leq$ 0.5 solar  and O/H solar.

The continuum SED of SGRB130603B (de Ugarte Postigo 2014) appears in Fig. 4. 
The star background contribution to SGRB130603B corresponds to \Ts=6000K.
For SGRB051221a the photometric data are too few to be used for modelling the SED.

\subsection{GRB111117A host SED  (Selsing et al 2017)}

Selsing et al (2017) presented spectroscopy and photometry data of  GRB 111117A host galaxy   with
an estimated  redshift of z=2.211.
The galaxy is actively forming stars. At z=2.211 GRB111117A is the most
distant high-confidence SGRB detected to date. 
Spectroscopic observations used VLT/X-shooter (Vernet et al 2011) at four separate epochs.
The burst was observed 38 hours after the BAT trigger.
Selsing et al determined the redshift value from the simultaneous detection of emission lines belonging
to \Ly, [OII]3727, [OIII]5007 and \Ha, but \Hb~  was detected at low significance.
 They used F$_{H\alpha}$=
4.1$\times$10$^{-17}$ \erg to calculate the SFR=18$\pm$3\msol yr$^{-1}$. No dust  was deduced from the \Ly,
so the spectra were  not reddening  corrected.
Selsing et al photometric data  appear in Fig. 5. They are  reproduced by a bb flux
corresponding to \Tbb=8000 K, in agreement with Margutti et al (2012)
who  reported that GRB111117A has a young age and vigorous star formation.

\section{Concluding remarks}

\begin{table*}
\centering
\caption{Results  of detailed modelling calculations}
\begin{tabular}{lcccccccccccccccc} \hline \hline
\  SGRB        & \Vs    &\n0  & $D$  & 12+log(N/H)& 12+log(O/H) & \Ts      &  $U$ & log(F) &\Hb   &T$_{bb}$  & z \\
\              & \kms   &\cm3 & pc   &     -      &      -      &  10$^4$K &  -   & $^1$   & $^2$ &   K       &         \\ \hline
\ NGC4993      & 100    &300  & 0.3  & 8.         &8.82         &   -      &  -   &  9.55  &0.05  &4000      &0.009873 \\
\ 050709       & 120    & 250 & 1.   & 7.25       &  8.81       & 4.8      &  0.06&-       &0.19  &3000      &0.167 \\
\ 051221a      & 150    &100  &0.07  &7.11        &8.82         &8.3       &0.007 &-       &0.007 &-         &0.546     \\ 
\ 061006       & 100    &100  & 0.27 & 7.0        &8.82         &5.7       &0.05  &  -     &0.006 &-         &0.4377   \\
\ 061210       & 100    & 300 & 0.93 & 6.9        &8.80         &4.3       &0.03  &-       &0.167 &-         &0.4095 \\
\ 070724       & 120    & 250 & 1.   & 7.25       &  8.81       & 4.9      &0.013 &-       &0.072 &-         &0.4571\\
\ 100206A      & 100    &120  &1.7   &8.04        &8.78         &4.4       &0.9   &-       &0.003 &2000      & 0.4068   \\
\ 100206A N    & 100    &250  &1.1   &8.25        &8.82         &3.6       &0.6   &-       &0.025 &-         &\\
\ 100206A C    & 100    &200  &1.1   &8.04        &8.82         &4.6       &0.6   &-       &0.0098&-         &\\
\ 100206A S    & 120    &200  &0.7   &8.15        &8.82         &4.        &0.6   &-       &0.0092&-         &\\
\ 130603B OT$^3$ & 140  &120  &1.8   &7.30        &8.82         &3.6       &0.014 &-       &0.014 &6000      &0.356     \\
\ 130603B OT$^4$ & 150  &130  &0.3   &7.30        &8.80         &3.5       &0.01  &-       &0.016 &-         &\\
\ 130603B core   & 150  &130  &0.3   &7.48        &8.80         &3.5       &0.01  &-       &0.016 &-         &\\
\ 130603B arm    & 140  &120  &0.3   &7.7         &8.82         &3.6       &0.014 &-       &0.03  &-         &\\
\ 130603B $^5$   & 120  &100  &1.    &7.47        &8.813        &3.8       &0.007 &-       &0.014 &-         &\\ \hline
\end{tabular}

$^1$F is in photon cm$^{-2}$ s$^{-1}$ eV$^{-1}$ at the Lyman limit; $^2$ in \erg calculated at the nebula;
$^3$ X- shooter; $^4$ FORS; $^5$DEIMOS

\end{table*}

In this work we have presented the analysis of line and continuum spectra of  a few SGRB host galaxies  
by exploiting the 
 spectroscopic and photometric data  available in the literature. 
For   some  objects included in our sample the spectra  show   the [OIII]/\Hb~ and [NII]/\Ha~ line ratios. 
These  diagnostics (e.g.  Baldwin et al 1981, Kewley et al 2001, etc)
 are generally used  to  recognise   the  type of flux which photoionizes  the GRB host gas 
(a bb  for   SBs, HII regions or a power-law for  AGNs).  
 Even if the literature data are sometimes  poor  and have not
always allowed us to  constrain the analysis results, we have nevertheless managed by the detailed 
modelling of   other line ratios  (e.g. [OII]3727+/\Hb, [SII]6717/[SII]6731, etc)
to obtain a more complete picture of the gas physical conditions (i.e. shock velocity, preshock density,
geometrical thickness of the emitting clouds, flux intensity from the AGN or from the SB, SB effective
temperature and ionization parameter) and chemical abundances.

We have investigated in particular NGC4993, the host galaxy of GW170817.
We confirm that an AGN, such as  a LLAGN or a LINER,   must be the source of     gas photoionization.
NGC4993 was formerly identified as a LLAGN on the basis of the BTP diagrams. 
Shock velocities ($\sim$100 \kms) and preshock densities ($\sim$300 \cm3), constrained by 
the simultaneous and consistent fit of the 
line ratios and of the SED, are similar to those found in the AGN NLR.  
By modelling the continuum SED  we have  also found 
that  the contribution of an old stellar  population background  with  \Tbb=4000K is  very high relative 
to the Bremsstrahlung in the near IR range  and    higher by a factor $>$ 100 than for  local  AGN and  SB  galaxies.  
O/H and N/H relative abundances   are  solar. 

 On the basis of the results obtained for NGC 4993 and for LGRB hosts (Contini 2018) and  for different
types of local galaxies (AGN SB, etc e.g. Contini \& Contini 2007) we suggest
that Bremsstrahlung  from the  clouds  which emit the line spectra could also  be
 the main contributor  to the continuum in the radio, optical,  UV and soft X-ray  range of SGRB hosts, while  
the stellar background bb radiation in the IR-optical domain is constrained by  
the photometry observations  in the IR.
Moreover,  dust reprocessed radiation  in the IR range (consistently calculated with the gas Bremsstrahlung) and
the flux from the photoionising source in the X-ray  are also considered. 
The  results calculated  for the SGRB hosts are summarized in Table 7.

In Fig. 6  diagrams  we have added the    N/O ratios calculated  for the  SGRB hosts as a function 
of the redshift to those  presented for
LGRB hosts, SN hosts of many types, HII regions, AGN, LINER and SB galaxies (Contini 2016). 
The right diagrams show the results for  O/H and N/H  as a  function of z. 
It is evident  that O/H is about constant at the solar value, while N/O covers a large range
of values determined by the N/H  ratios  (6.8$\leq$log(N/H)+12$\leq$8.25)
throughout a very small redshift range centered  around z$\sim$0.4.
N/O abundance ratios calculated in  GRB100206B along different positions -- observed by Perley et al.--
reach  the  maximum values calculated for LGRB hosts at higher redshifts. 
At present, only  few data at other z are available. 
NGC4993 is located within the AGN and LINER region at a redshift close to local.
Compared with LGRB and SGRB hosts  (Contini 2016), the most significant  characteristic of NGC4993
is the presence of an   AGN source.
Actually, NGC4993  is the only SGRB host lying among AGN, LINERs and LLAGN at low z (see Fig. 6).
Whether this is an isolated case  it is not yet known.
However, the  high evolution age and/or  a different nuclear development related to  the  primary
 nitrogen should be considered.  More data are needed. 

Most of the  other SGRB are located    
at the confluence (at z close to 0.4) of the N/O decreasing slope with decreasing z of  GRB hosts  
and of  the increasing N/O slope with decreasing z of SN hosts at lower z (Fig. 6). 
Intermediate-mass stars between 4 and 8 \msol dominate nitrogen production, which is primary at low metallicities,
but when 12+log(O/H) exceeds 8.3, secondary N production occurs, increasing at a faster rate than O 
(Henry et al 2000). The data are few, but it seems that N/O ratios in the present SGRB sample follow a different trend.

\section*{Acknowledgements}
I am very grateful to the referee for critical suggestions which substantially
improved the presentation of the paper.

\section*{References}

\def\ref{\par\noindent\hangindent 18pt}

\ref Baldwin, J., Phillips, M., Terlevich, R. 1981, PASP, 93, 5
\ref Barthelmy, S.D. et al 2005, Nature, 438, 994
\ref Bell, A.R. 1977, MNRAS, 179, 573
\ref Berger, E. 2014, ARA\&A, 52, 43
\ref Berger, E. 2013, ApJ, 765, 121 
\ref Berger, E. 2009, ApJ, 690, 231
\ref Berger, E. et al 2005, Nature, 438, 988
\ref Blanchard, P.K. et al 2017, ApJL, 848, L22
\ref Bloom, J.S., Kulkarni, S.R., Djorgavski, S.G. 2002, AJ, 123, 1111
\ref Contini, M. 2018, submitted
\ref Contini, M. 2017a, MNRAS, 469,3125
\ref Contini, M. 2017b, MNRAS, 466, 2787
\ref Contini, M. 2016, MNRAS, 460, 3232
\ref Contini, M. 2004, MNRAS, 675, 683
\ref Contini, M. \& Contini, T. 2007, AN, 328, 953
\ref Contini, M. \& Viegas, S.M. 2001a ApJS, 137, 75
\ref Contini, M. \& Viegas, S.M. 2001b ApJS, 132, 211
\ref Contini, M. \& Aldrovandi, S.M.V. 1983 A\&A, 127, 15
\ref Contini, M., Prieto, M.A. \& Viegas, S.M. 1998a, ApJ, 505, 621
\ref Contini, M., Prieto, M.A. \& Viegas, S.M. 1998b, ApJ, 492, 511
\ref Cox, D.P. 1972, ApJ, 178, 143
\ref Cucchiara, A. et al 2013, ApJ, 777, 94
\ref D'Avanzo, P. 2015, JHEAp, 7, 73
\ref de Ugarte Postigo, A. et al. 2014, A\&A, 563, 62 
\ref De Voucouleurs, G., De Voucouleurs, A., Corwin Jr., H.J., Buta, R.J., Paturel, G., Fouque, P. 1991
Third reference catalogue of bright galaxies, RC3.9
\ref Draine, B.T. \& Lee, M.M. 1994, ApJ, 285, 89
\ref Eichler, D., Livio, M., Piran, T., Schramm, D.N. 1989, Nature, 340, 126
\ref Ferland G. et al 1995,  The Analysis of Emission Lines: A Meeting
in Honor of the 70th Birthdays of D. E. Osterbrock \& M. J. Seaton,
proceedings of the Space Telescope Science
Institute Symposium, held in Baltimore, Maryland May 16--18, 1994, 
Eds.: Robert Williams and Mario Livio, Cambridge University Press, p. 83, ArXiv:160308902
\ref Fong, W. et al 2013, ApJ 769, 56
\ref Fox, D.B., Frail, D.A., Hurley, K.C. et al 2005, Nature, 437, 845
\ref Fruchter, , A.S et al 2006, Nature, 441, 463
\ref Gehrels, N.,White, N., Barthelmy, S. et al 2004, ApJ, 611, 1005
\ref Grevesse, N., Sauval, A.J. 1998 SSRv, 85, 161
\ref Haggard, D. et al 2017, ApJ, 848, L25
\ref Henry, R.B.C., Edmund, M.G.\& K\"{o}ppen, J. 2000, ApJ, 541, 660
\ref Hunt, L.K. et al 2014, A\&A, 565, A112
\ref Kann, D.A. 2011, ApJ, 734, 96
\ref Kauffmann, G. et al 2003, MNRAS, 346, 1055
\ref Kennicutt, R.C. 1998, ARA\&A, 36, 189
\ref Kewley, L., Dopita, M.M., Sutherland, R., Heisler, C., Trevena, J. 2001, ApJ, 556, 121
\ref Kouvelioutou, C et al 1993, ApJL, 413, L101
\ref Lambert, A. \& Valentijn, E.A. 1989, The Surface Photometry Catalogue of the ESO-Uppsala, 1989,
Garching bei Munchen:ESO 
\ref Levan, A.J. et al. 2017 ApJL (in press. arXiv:1710.05444)
\ref Margutti, R., Berger, E. , Fong, W. et al 2012, ApJ, 756, 63
\ref Maraston, C. 2005, MNRAS, 362, 799
\ref Niino, Y. et al 2016 Publ. Astron. Soc. Japan,  ArXiv:1606.01983
\ref Osterbrock, D.E. 1974, in Astrophysics of Gaseous Nebulae, W.H Freeman and Co., San Francisco
\ref Palmese, A. et al 2017, ArXiv:1710.06748
\ref Pan, Y.-C. et al 2017, ApJ, 848, L30
\ref Perley, D.A. et al. 2012, ApJ, 758, 122
\ref Rigby, J.R.\& Rieke, G.H. 2004, ApJ, 606,237
\ref Rosswog, S., Ramirez-Ruiz, E., Davies, Melvyn B, 2003, MNRAS, 345, 1077
\ref Savaglio, S., Glazerbrook, K., Le Borgne, D. 2009, ApJ, 691, 182
\ref Selsing, J. et al 2017 arXiv:1707.01452
\ref Soderberg, A.M. et al 2006, ApJ, 650, 261
\ref Vernet, J. et al.   2011, A\&A, 536, A105
\ref Viegas, S.M. \& Contini, M. 1994, ApJ, 428, 113
\ref Villar, V.A. et al 2017, arXiv:1710.11576
\ref Williams, R.E. 1967, ApJ, 147, 556
\ref Wu, Q., Feng, J., Fan, X. 2017, arXiv:1710.09590

\end{document}